\documentclass{JINST}
\usepackage{subfig}

\title{Simulation of VUV electroluminescence in micropattern gaseous 
detectors: the case of GEM and MHSP}

\author{C. A. B. Oliveira$^{ab}$\thanks{Corresponding
author.}~, P. M. M. Correia$^a$, H. Schindler$^c$, A. L. Ferreira$^a$, C. M. B. Monteiro$^d$, J. M. F. dos Santos$^d$, S. Biagi$^e$, R.
Veenhof$^c$, J. F. C. A. Veloso$^a$
\\
\llap{$^a$}i3N, Physics Department, University of Aveiro\\
  Campus de Santiago, 3810-193 Aveiro, Portugal\\
\llap{$^b$}Lawrence Berkeley National Laboratory\\
	One Cyclotron Road, Berkeley, 94720, CA, USA\\
  \llap{$^c$}European Organization for Nuclear Research CERN,\
  \\CH-1211, Gen\`{e}ve 23, Switzerland\\
  \llap{$^d$}University of Coimbra, Physics Department, Instrumentation Center\\
  3004-516 Coimbra, Portugal \\
  \llap{$^e$}University of Liverpool, Physics Department\\
  Liverpool L69 7ZE, United Kingdom \\
E-mail: \email{cabdoliveira@lbl.gov} \\}

\abstract{Electroluminescence produced   during   avalanche   development   in   gaseous   avalanche   detectors   is   an   useful 
information for triggering, calorimetry and tracking in gaseous detectors. Noble gases 
present high electroluminescence yields, emitting mainly in the VUV region. The photons can provide signal readout if appropriate photosensors are used. 

Micropattern gaseous detectors are good candidates for signal amplification in high background and/or low rate experiments due to their high electroluminescence yields and radiopurity.
In this work, the VUV light responses of the Gas Electron Multiplier and of the Micro-Hole \& Strip Plate, working with pure xenon, are 
simulated   and   studied   in   detail   using   a   new   and   versatile   C++ toolkit.  
It is shown that the solid angle subtended by a photosensor placed below the microstructures depends on the operating conditions. The   obtained  absolute EL   yields, 
determined for different gas pressures and as functions of the applied voltage, are compared with those determined experimentally.}


\begin{document}

\section{Introduction}

Cryogenic dual-phase liquid/gas as well as gaseous detectors have been or are currently being developed for direct detection of WIMPs (Dark Matter candidates)~\cite{xenon100,arDM,Benetti_warp,lux,zeplin} and for neutrino less double beta decay (0$\nu\beta\beta$) searches ~\cite{exo,NEXT_CDR,NEXT_TDR}. These detectors consist of Time Projection Chambers (TPCs) operating with pure noble gases, where both primary VUV scintillation and primary ionization are produced by radiation interaction. The primary VUV light is used for \textit{start-of-event} triggering while the electrons resulting from ionization are driven to a region where their signal is amplified. It is important to have the highest possible signal gain, given the low rate and/or high background nature of these experiments. Because of this, electroluminescence (EL) is usualy used for signal amplification, rather than charge multiplication.
In this process, the electrons are accelerated by a suitable uniform electric field that, between collisions, supplies them with enough energy to excite but not to ionize atoms of the gas. For atmospheric pressure or higher, the excited atoms decay through the formation of excimers~\cite{Tanaka}. In the case of xenon, the result is the emission of VUV photons with a wavelength centered at 173 nm and with a Full-Width-at-Half-Maximum (FWHM) of 14 nm~\cite{Suzuki}. Further details about the electroluminescence process in pure noble gases can be found in reference~\cite{Oliveira_thesis} and references therein.

The future ton-scale Dark Matter and 0$\nu\beta\beta$ experiments will introduce new challenges in the design of large-volume TPCs~\cite{DARWIN}. Namely, the increase in the number of the commonly used PMTs for VUV light readout would lead to undesired high background levels. The high radiopurity, simplicity, low cost and the possibility of constructing large areas, make of Micropattern Gaseous Detectors (MPGDs) good candidates for the amplification of the primary ionization signal~\cite{THGEM_DM,GEM_dualphase,Buzu_advances}. In these devices, during avalanche multiplication, also VUV light is produced since both excitations and ionizations are induced by direct impact of electrons. It has been demonstrated that higher light gains than those obtained with parallel gaps can be achieved by using much lower voltages~\cite{cristina_GEM_MHSP}. The VUV EL produced during the avalanches can be detected, e.g, by Geiger mode avalanche photodiodes (GM-APDs)~\cite{apd_vuv_1,apd_vuv_2}, also characterized by a negligible natural radioactivity~\cite{apd_radiopure}.

The simulation of EL produced in MPGDs is thus of major importance for the correct design of future detectors. In reference~\cite{cristina_GEM_MHSP} it was determined experimentally the absolute VUV EL yields of the Gas Electron Multiplier (GEM) and of the Micro-Hole \& Strip Plate (MHSP). Recently, the yield of the thicker version of the GEM (the THGEM) was also measured~\cite{cristina_GEM_THGEM}. In this work, we use a recent developed and flexible Monte Carlo simulation toolkit~\cite{OliveirauE} to estimate the properties of VUV EL produced in GEM and MHSP. The obtained EL yields are compared, for the case of pure xenon at different pressures, with the measurements.

\section{Simulation}

The toolkit used in this work is implemented in C++ and based on the Garfield~\cite{garfield} and Magboltz 8.9.7~\cite{magboltz1,magboltz2} programs. Through this flexible toolkit it is possible to follow the electrons through the gas in nearly arbitrary field geometries. Electrons are tracked
at the microscopic level (atomic/molecular) using Monte Carlo procedures and cross-sections available in Magboltz. In the case of xenon, the gas studied in this work, beyond cross sections for elastic collisions and ionizations, the program parametrizes the excited energy levels as functions of 50
energy groups. It is possible to obtain information about each excitation produced in the gas, namely: spatial position, time, atomic level and spent energy. Additional details about the toolkit can be found at references~\cite{Oliveira_thesis,OliveirauE}.

We assume that every excited atom gives rise to the isotropic emission of a VUV photon~\cite{Tanaka,Oliveira_thesis,3d}. The toolkit and the model were previously validated for uniform electric field geometry in reference \cite{OliveirauE} by comparing the simulated results with measurements~\cite{expy_xe_ue,expy_ar_ue} and earlier Monte Carlo work~\cite{3d}.

Electric fields created by microstructures like GEM and MHSP can not be determined analytically due to their complex shape and to the presence of insulators. Thus, field maps were constructed using ANSYS12$\textsuperscript{\textregistered}$~\cite{ansys}, a Finite Element Method (FEM)~\cite{fem} program. The \textit{3-D 10-Node Tetrahedral Electrostatic Solid} finite element was used since it is suitable for modeling irregular meshes that need to adapt to curved edges. Since the hole and electrode patterns of the GEM and MHSP are periodic, we constructed unitary cells, being the longitudinal axis of the holes parallel to the $zz'$ axis. Mirror symmetries applied in the $xx'$ and $yy'$ directions were used, creating maps of ideal microstructures with infinite areas.

\section{GEM}

\subsection{Simulation details}

For the simulations, a standard GEM was considered, consisting of a 50 $\mu$m thickness Kapton$\textsuperscript{\textregistered}$ foil covered with 5 $\mu$m thickness copper layers on both sides and perforated with bi-conical holes of 50 and 70 $\mu$m diameter at the inner and outer apertures respectively, arranged in a regular hexagonal layout with an edge of 140 $\mu$m length.

In reference~\cite{cristina_GEM_MHSP}, the primary charge was generated using a collimated (1 mm radius) 22.1 keV X-ray beam and the VUV EL photons were measured by a Large Area Avalanche Photodiode (LAAPD) with an active diameter of 16 mm placed below the GEM. The drift and induction fields were 0.5 and -0.1 kVcm$^{-1}$ respectively. The unusual reversed induction field was used experimentally so that no charge readout below the GEM was needed and that the LAAPD could replace it. This experimental field setup was implemented in the simulations, as well as the beam characteristics.

The primary electrons were released isotropically 250 $\mu$m above the GEM (measured in the $zz'$ direction), in a region where the equipotential surfaces are completely flat and the electric field is uniform. The $\left(x,y\right)$ coordinates of the starting point were sampled randomly according to the geometry of the X-ray beam, the corresponding absorption probability along the 8 mm length drift region and the transversal diffusion that the primary electrons undergo before arriving at the $z = 250$~$\mu$m plane. Each electron started with a kinetic energy sampled randomly according to the energy distribution given by Magboltz for the experimental drift field. A set of 3$,$000 primary electrons was simulated.
Pure xenon at a temperature of 300~K and pressures of 1.0, 1.5 and 2.5 bar was considered. 

\subsection{VUV EL yield}

\begin{figure}[h!]
\centering
\includegraphics[width=0.75\textwidth]{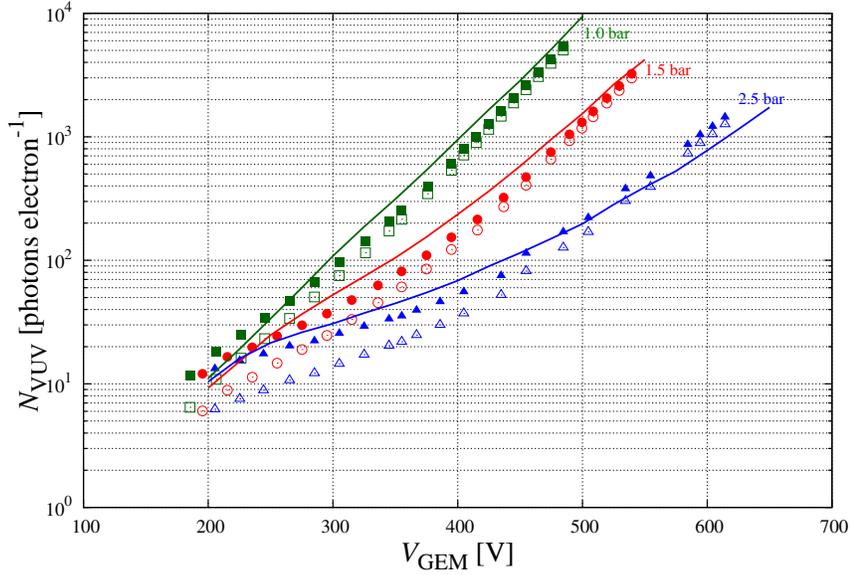}
\caption{Average number of VUV photons produced per primary electron as a function of the potential applied between the top and the bottom electrodes of the GEM, shown for different pressures of xenon at 300~K. The simulated results (lines) are compared with the total VUV EL yield reported in~\cite{cristina_GEM_MHSP} (open symbols). The latter were corrected for the actual fraction of solid angle (see Section 3.3 for further details) and the result is shown as closed symbols.}
\label{fig:gem_y}
\end{figure}

The calculated absolute VUV EL yield, i.e., the total number of VUV photons produced per primary electron, is shown in Figure~\ref{fig:gem_y} (lines) as a function of the potential applied between the top and the bottom electrodes of the GEM, $V_\mathrm{GEM}$. Results for the different simulated pressures are compared with those of the measurements reported in reference~\cite{cristina_GEM_MHSP} (open symbols).
Specially for high voltages, the simulated VUV EL yield approaches the experimental results (which have an associated uncertainty of the order of 20 \%). However, some differences between the data are present.

\subsection{Solid angle}
\label{subsec:solid_angle}
\begin{figure}[h!]
  \centering
  \subfloat[$V_\mathrm{GEM} = 200\textrm{ V}$]{\label{fig:hz_200V}\includegraphics[width=0.45\textwidth]{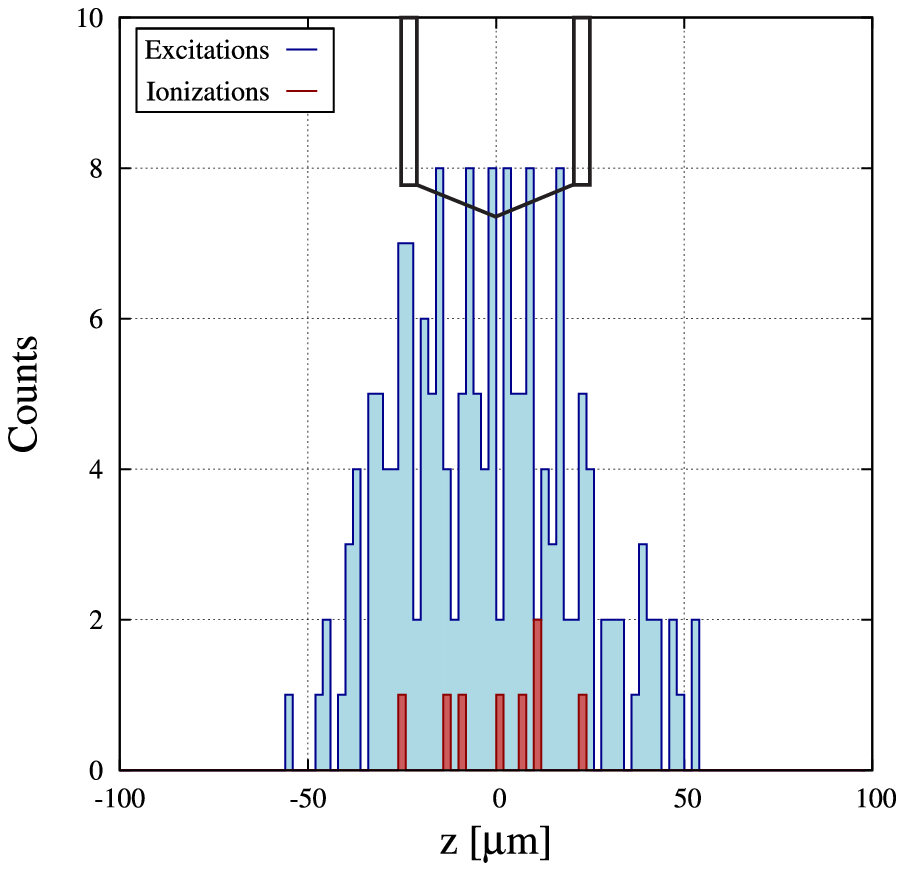}}\hspace{0.2cm}
  \subfloat[$V_\mathrm{GEM} = 300\textrm{ V}$]{\label{fig:hz_300V}\includegraphics[width=0.45\textwidth]{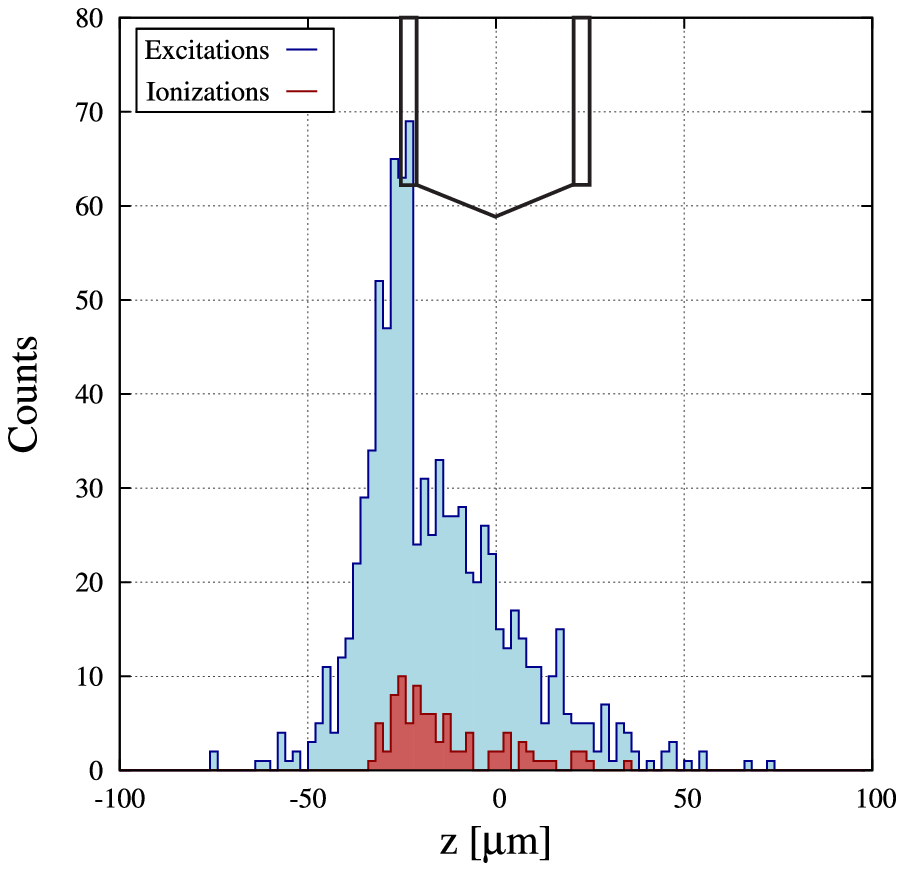}}\\
  \subfloat[$V_\mathrm{GEM} = 400\textrm{ V}$]{\label{fig:hz_400V}\includegraphics[width=0.45\textwidth]{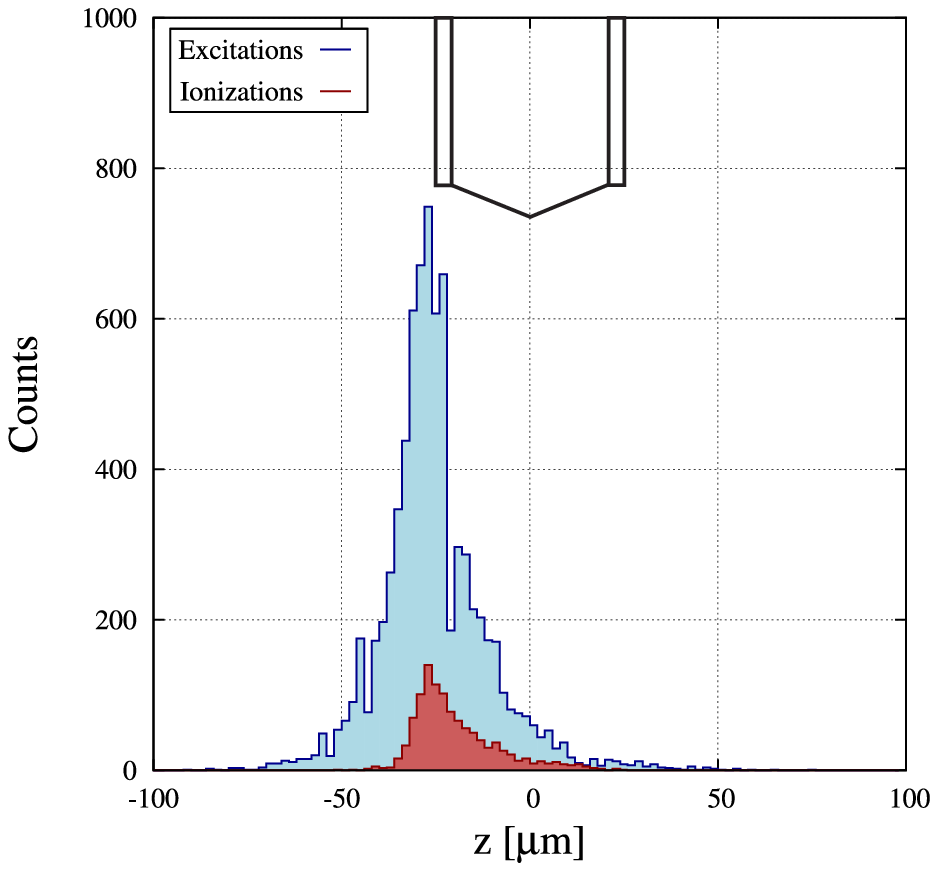}}\hspace{0.2cm}
  \subfloat[$V_\mathrm{GEM} = 500\textrm{ V}$]{\label{fig:hz_500V}\includegraphics[width=0.45\textwidth]{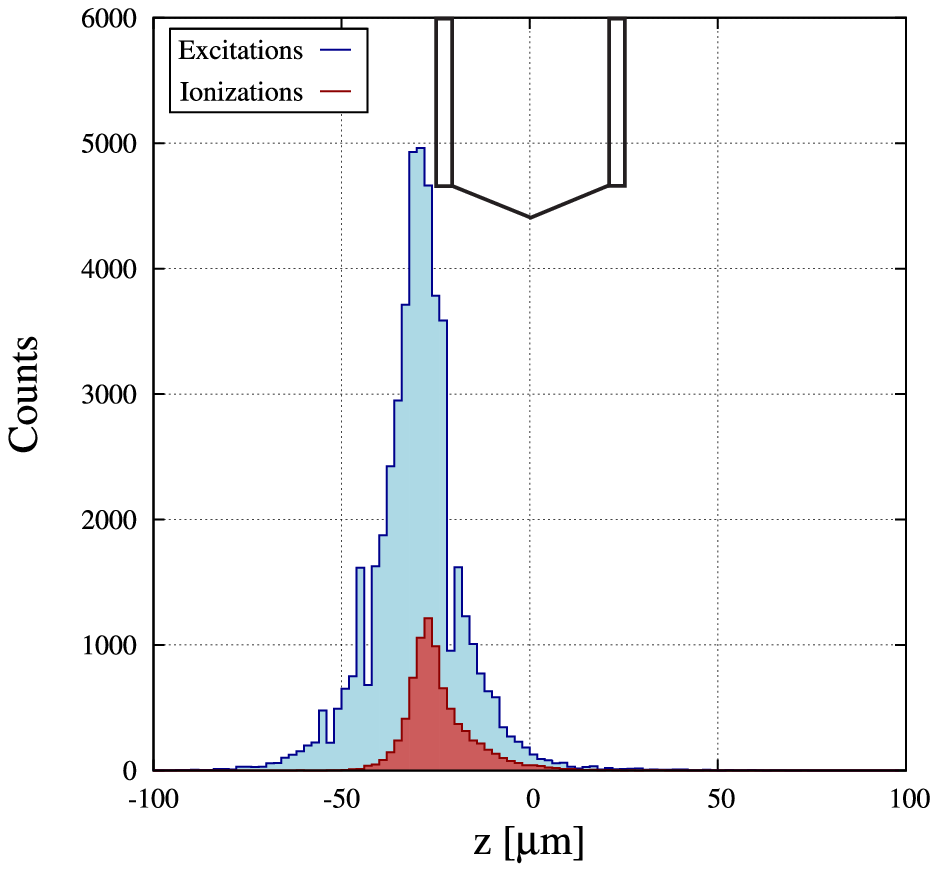}}
  \caption{Distribution of the $z$ coordinate of the points where excitations and ionizations occurred, shown for some voltages applied along the GEM's holes. For each case, a set of 10 primary electrons was simulated for xenon at 1.0 bar and 300 K. The edges of the microstructure are proportionally represented at each plot. The electrons enter the GEM hole for positive values of $z$ and drift toward negative values.}
  \label{fig:hz}
\end{figure}

Not all of the VUV photons produced in the gas impinge the surface of the photodetector. The LAAPD has a subtended solid angle that need to be considered in the calibrations performed during the measurements. In reference~\cite{cristina_GEM_MHSP} this solid angle was calculated considering that all the VUV photons were emitted from a plane corresponding to the bottom surface of the bottom electrode, being independent of the applied voltage and of the gas pressure.

The plots of Figure~\ref{fig:hz} show the distribution of the $z$ coordinate of the points where excitations occurred, for some applied potentials and for xenon at 1.0 bar and 300 K. In the same plots, it is included the same distribution for the ionization points. For a fixed pressure, at low voltages, the number of produced secondary charges is low (refer also to Figure~\ref{fig:ne}) and the microstructure works almost only in EL mode, being ionizations rare. The regions where more atoms are excited reflect the volumes where higher electric fields exist. The fraction of excitations and thus, of VUV photons emitted in the upper half part of the holes is  roughly  the same as that in the lower half part. As the applied voltage increases, the size of the avalanche increases and a higher fraction of secondary electrons is created in its last stage (close to the bottom electrode), reflecting the cascade nature of avalanches. These secondary charges also excite atoms and thus induce the emission of VUV photons. The fraction of photons emitted in the upper half part of the holes is now much lower than that of photons originated in the lower half part.  Since the GEM itself blocks a higher fraction of photons originated in the upper regions of the holes than of those from the lower region, the subtended solid angle increases with the applied voltage.

For a given applied voltage, lower pressures mean higher gains, due to the increase of the energy achieved by electrons between collisions. The ratio between the number of photons originated from the lower region of the holes and the number of photons coming from the upper region is, thus, higher for lower pressures. The subtended solid angle increases as the gas pressure decreases.

In conclusion, the fraction of solid angle subtended by the LAAPD (or any other photosensor placed below the GEM), $\Omega$, is dependent on the gas pressure and on the applied voltage. Experimentaly it is difficult to determine $\Omega$ as function of these two quantities and thus, in reference~\cite{cristina_GEM_MHSP} it was considered as being the same for all the conditions, giving more importance to photons produced in the lower region of the holes, which is appropriated only for high VUV EL yields. 

\begin{figure}[h]
\centering
\includegraphics[width=0.75\textwidth]{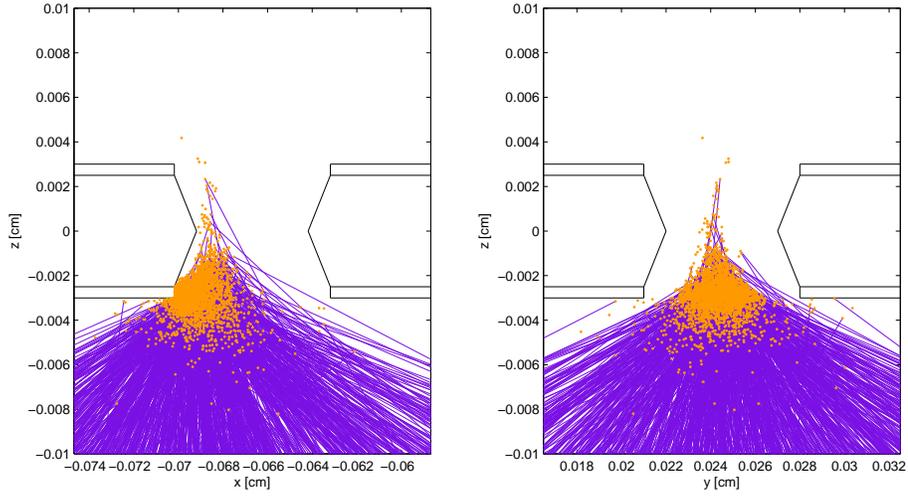}
\caption{Points where excitations occured during one avalanche happening in Xe at 1.0 bar and 300~K and for $V_\textrm{GEM}=500$ V. The trajectories of the VUV photons that hit the LAAPD surface are also included. The edges of the microstructure are proportionally represented at each plot.}
\label{fig:gem_ilustration}
\end{figure}

Using the information about the positions where excitations occurred, available through the simulation toolkit, it was possible to estimate the solid angle for each of the simulated conditions, considering that a VUV photon is emitted isotropically from each excitation point. According to the detector geometry, the photon can impinge the photosensor surface, be blocked by the microstructure or hit the detector walls. In the latter case no reflection is considered. The value of $\Omega$ is determined as being the ratio between the number of photons that impinge the photosensor surface and the total number of photons produced in the GEM. For ilustration, in Figure~\ref{fig:gem_ilustration} are represented the points where excitations occured during one avalanche happening for a gas pressure of 1.0 bar and $V_\textrm{GEM}=500$ V. The trajectories of the VUV photons that hit the LAAPD surface are also included.

\begin{figure}[h!]
\centering
  \subfloat[]{\label{fig:omega_v}\includegraphics[width=0.49\textwidth]{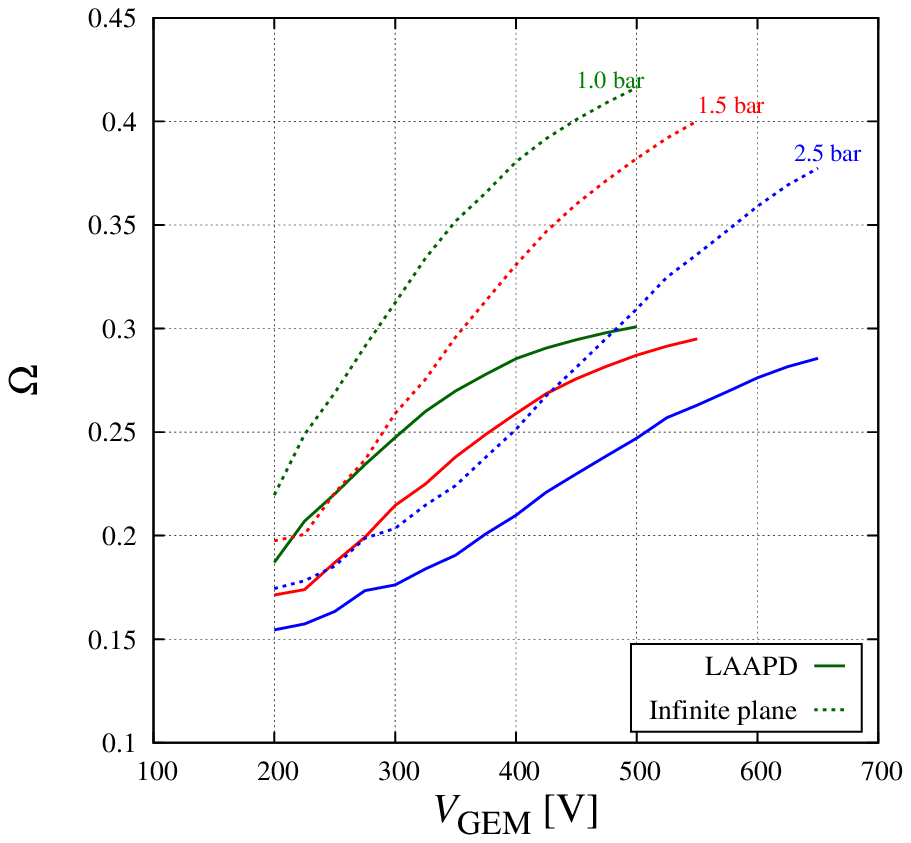}}\hspace{0.1cm}
  \subfloat[]{\label{fig:omega_gain}\includegraphics[width=0.49\textwidth]{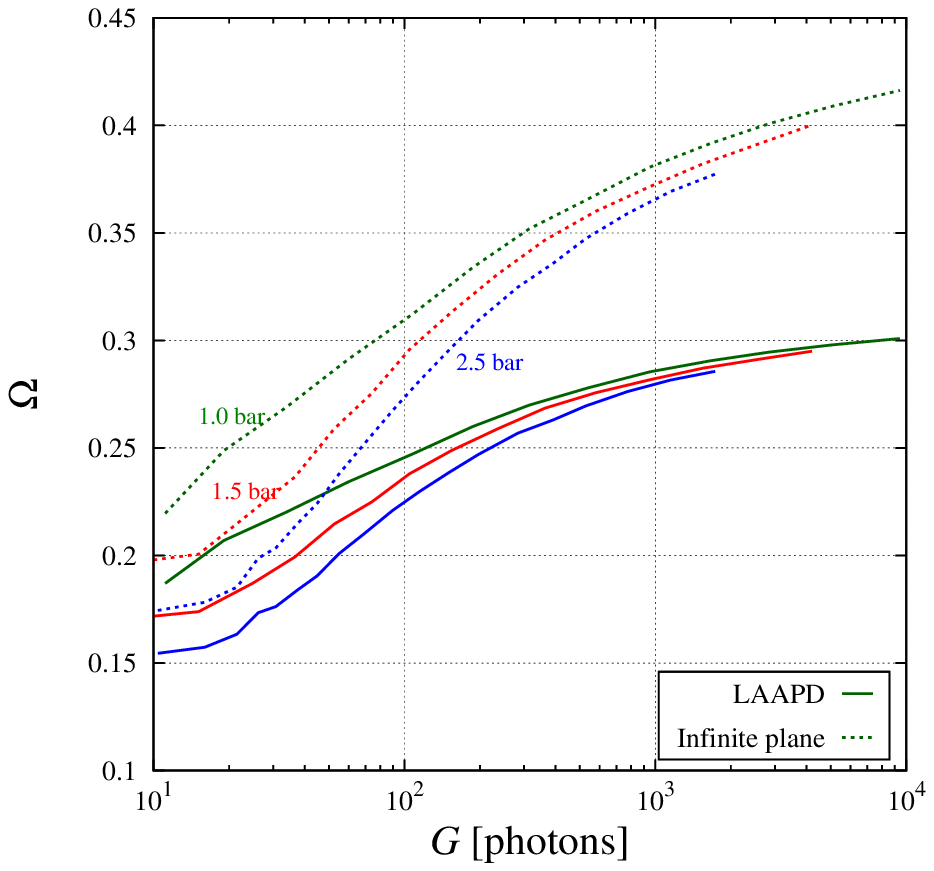}}
\caption{Fraction of solid angle subtended by the LAAPD (full lines) as function of $V_\mathrm{GEM}$ (left) and the absolute VUV yield (right). Curves for the different simulated pressures are shown. The plots also show the estimated values of the solid angle subtended by an infinit plane located below the GEM (dashed lines).}
\label{fig:gem_omega}
\end{figure}

Figure~\ref{fig:omega_v} shows the calculated $\Omega$ as a function of $V_\mathrm{GEM}$, for the different considered pressures. The estimated values of the solid angle subtended by an infinit plane located below the GEM are also shown. These values represent the maximum light detection coverage that can be achieved. In Figure~\ref{fig:omega_gain}, the values of $\Omega$ are shown as functions of the light gain, i.e., the absolute VUV yield. For higher gains, $\Omega$ starts to saturate since almost all the excitations happen at the exit of the hole, reason why $\Omega$ also approaches the value calculated in reference~\cite{cristina_GEM_MHSP}, $\Omega=0.32$.

The experimental values of the absolute VUV yield, corresponding in Figure~\ref{fig:gem_y} to the open symbols, were corrected for the actual fraction of solid angle. The corrected values are shown in the same figure as closed symbols. It is clearly visible the better agreement between corrected experimental results and Monte Carlo simulations, specially for lower voltages. The overall agreement is remarkable considering the uncertainty in the measured values ($\sim$20 \%). The small remaining differences can be attributed to the effect of the charging-up of the insulator surface, not considered in the simulations; to intrinsic errors in the electric field calculations, characteristic of the FEM method; to imperfections in the GEM geometry introduced during its construction; and to the uncertainty in the temperature at which the measurements were performed. However, the contribution of these effects seems to be small, given the achieved good agreement.

\subsection{Ratio between light and charge carriers}

\begin{figure}[b]
\centering
  \subfloat[]{\label{fig:ratio}\includegraphics[width=0.49\textwidth]{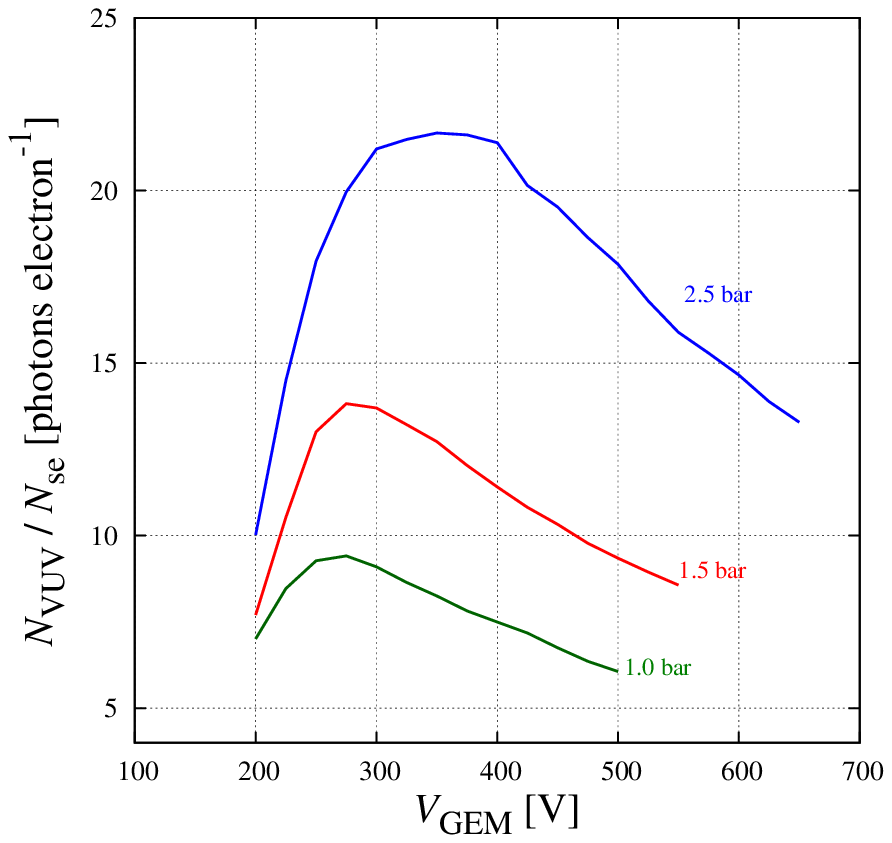}}\hspace{0.1cm}
  \subfloat[]{\label{fig:ne}\includegraphics[width=0.49\textwidth]{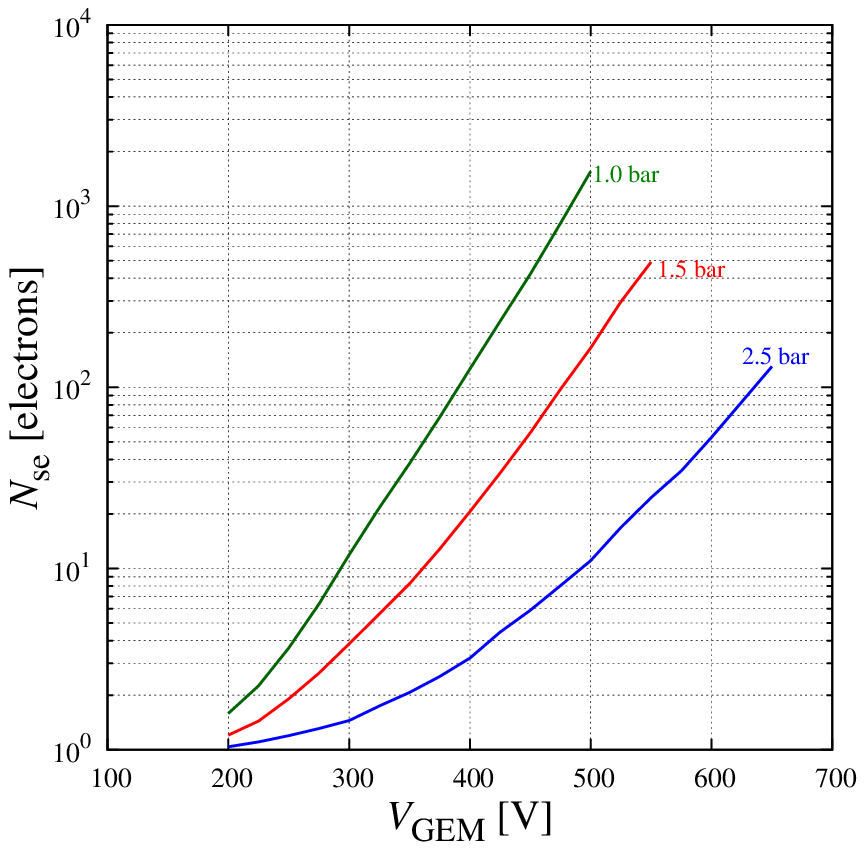}}
\caption{\textbf{a)} Ratio between the number of excitations  and the number of ionizations produced during avalanches developed in a GEM, shown as a function of the voltage applied between the top and the bottom electrodes, $V_\mathrm{GEM}$, and for different pressures of xenon at 300~K. \textbf{b)} Average number of secondary electrons produced per primary electron (including the latter) for the same conditions.}
\label{fig:ratio_ne}
\end{figure}

The spatial distributions presented in Figure~\ref{fig:hz} show that, in a GEM, the number of excitations produced during avalanches is much higher than the number of ionizations. The ratio between these two numbers is shown, as a function of $V_\mathrm{GEM}$ and for the different simulated xenon pressures, in Figure~\ref{fig:ratio}. As reference, in Figure~\ref{fig:ne} it is shown the average number of secondary electrons produced per primary electron, for the same conditions.

For the same applied voltage, as higher  pressures are considered, the maximum energy acquired by electrons between collisions decreases, since their free path is lower. Since the energy threshold for ionizations is higher than for excitations, the latter are more likely for higher pressures. Considering the same pressure and starting with low voltages ($V=200$ V), the ratio initially grows since the electric field is increasing and the maximum kinetic energy reached by the electrons is enough for excitations but barely for ionizations. For a certain voltage, at some of the collisions, electrons can achieve an energy slightly higher than the ionization threshold.  At these energies, the ionization cross section is higher than that for the sum of the excitations and thus the ratio starts to drop.

\section{MHSP}

\subsection{Simulation details}

\begin{figure}[b]
\centering
\includegraphics[width=0.72\textwidth]{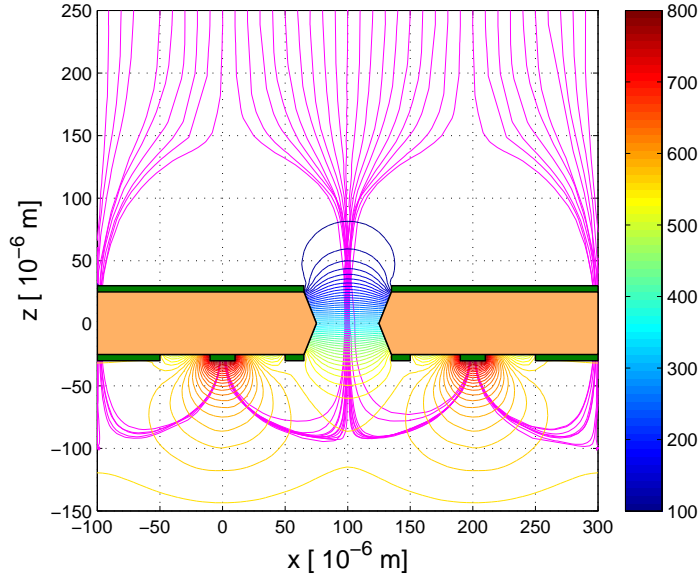}
\caption{Equipotential lines produced in the MHSP microstructure when $V_\mathrm{CT}=450$ V and $V_\mathrm{AC}=250$ V are applied. Isolines are shown for potentials between 100 V and 800 V in steps of 10 V, according to the colorbar shown at the right. The edges of the microstructure are proportionally represented at the same plot. The view plane, which contains the page, is $y=35\textrm{ }\mu$m and contains the center of the represented hole. The magenta lines show the average paths travelled by electrons (drift lines) starting 250 $\mu$m above the microstructure, being the $x$ coordinate equally distributed along an horizontal line that is contained in the view plane. Drift lines beginning at $x<0\textrm{ }\mu$m and at $x>200\textrm{ }\mu$m are focused into the holes which centers are contained in a plane $70\textrm{ }\mu$m behind the view plane.}
\label{fig:mhsp_v}
\end{figure}

We simulated also the VUV EL response of a standard MHSP, consisting of a 50 $\mu$m thickness Kapton$\textsuperscript{\textregistered}$ foil covered with 5 $\mu$m thickness copper layers on both sides and perforated with bi-conical holes of 50 and 70 $\mu$m diameter at the inner and outer apertures respectively, arranged in an hexagonal pattern. The hole-pitch was 200 $\mu$m along the $xx'$ direction (perpendicular to the strips) and 140 $\mu$m along the $yy'$ direction (parallel to the strips). The width of the cathodes and anodes was 100 and 20 $\mu$m respectively. The shape of the microstructure is depicted in Figure~\ref{fig:mhsp_v}. The detector geometry of reference~\cite{cristina_GEM_MHSP} was considered. There, the primary charge was extracted from a semi-transparent CsI photocathode by photons emitted from a UV lamp. The VUV EL photons produced during avalanches extracted photoelectrons from a reflective CsI photocathode. The electrons were then collected by a metal mesh, being the induced current proportional to the number of produced VUV photons. The drift and induction fields were 0.1 and -0.1 kVcm$^{-1}$ respectively.

The primary electrons were released isotropically 250 $\mu$m above the MHSP (measured in the $zz'$ direction), in a region where the equipotential surfaces are completely flat and the electric field is uniform. The $\left(x,y\right)$ coordinates of the starting point were sampled randomly according to an uniform distribution along the active area of the microstructure. Each electron started with a kinetic energy sampled randomly according to the energy distribution given by Magboltz for the drift field. A set of 500 primary electrons was simulated.
Pure xenon at 1.0 bar and 300 K was considered. 

\subsection{VUV EL yield}

\begin{figure}[b!]
\centering
\includegraphics[width=0.75\textwidth]{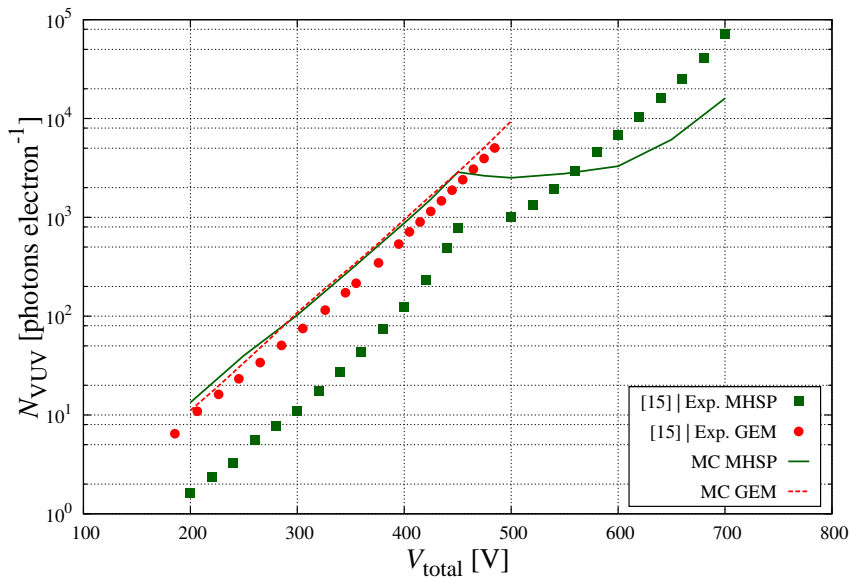}
\caption{Average number of VUV photons produced per primary electron in a MHSP as a function of the total applied voltage ($V_\mathrm{total}=V_\mathrm{CT}+V_\mathrm{AC}$). The results are shown for xenon at 1.0 bar and 300~K. Those obtained through the simulation (green line) are compared with the total VUV EL yield reported in~\cite{cristina_GEM_MHSP} (filled squares). The results of simulations and measurements (uncorrected for the variation of $\Omega$) for the case of GEM are also included in the plot.}
\label{fig:mhsp_y}
\end{figure}

The calculated total VUV EL yield is shown in Figure~\ref{fig:mhsp_y} as a function of the total voltage, $V_\mathrm{total}$. This quantity corresponds to the sum of the potential applied between the cathode and the top electrodes, $V_\mathrm{CT}$, with the potential applied between the anode and the cathode, $V_\mathrm{AC}$. The results are compared with the measurements. The simulated and the experimental yields obtained for the case of GEM are also included in the plot. The two exponential regimes typical of the MHSP are clearly visible in the Monte Carlo curve: the first corresponds to the increase on $V_\mathrm{CT}$ while $V_\mathrm{AC}=0$ V, and the second to an increasing $V_\mathrm{AC}$ while the value of  $V_\mathrm{CT}$ is kept constant.

The Monte Carlo absolute VUV EL yield doesn't agree with the measured one. However, it is close to the experimental and simulated GEM's yields for the points where $V_\mathrm{AC}=0$ V (the initial exponential regime). Both the GEM and the MHSP considered in this work have exactly the same hole shape and dimensions, having the induction fields in both cases the same orientation. Thus, for the case of $V_\mathrm{AC}=0$ V, the VUV EL yields of the MHSP and the GEM should be very similar, which is not the case for the measurements. 
Experimentaly, for the case of the MHSP, the gas purification was achieved by diffusion, unlike as in the case of the GEM, for which it was achieved by convection. The xenon purity was thus affected, being possible that the measurement underestimated the VUV EL yield of the MHSP. It is known that small amounts of impurities have a big quenching effect in the production of VUV EL ligh in pure noble gases~\cite{expy_xe_ue}. 

For the second regime ($V_\mathrm{AC} > 0$ V), the experimental yield is now higher than the simulated one. This suggests imperfections in the construction of the microstructure, which is of peculiar difficulty. Namely, variations in the shape and dimensions of the anode have an important impact on the overall gain, both in terms of charge and light. Figure~\ref{fig:mhsp_v} shows the equipotential lines produced in the MHSP when $V_\mathrm{CT}=450$ V and $V_\mathrm{AC}=250$ V are applied, as well as the drift lines of some electrons. As shown, the highest electric fields exist near the anode, where the density of drift lines is also higher. Here, the drifting electrons have the highest probability of producing new ionizations. Besides that, this region consists of the second stage of charge amplification and thus, small variations in the field shape or amplitude have big effects on the overall gain.

\subsection{Spatial distribution of excitations}

\begin{figure}[h!]
\centering
\includegraphics[width=0.6\textwidth]{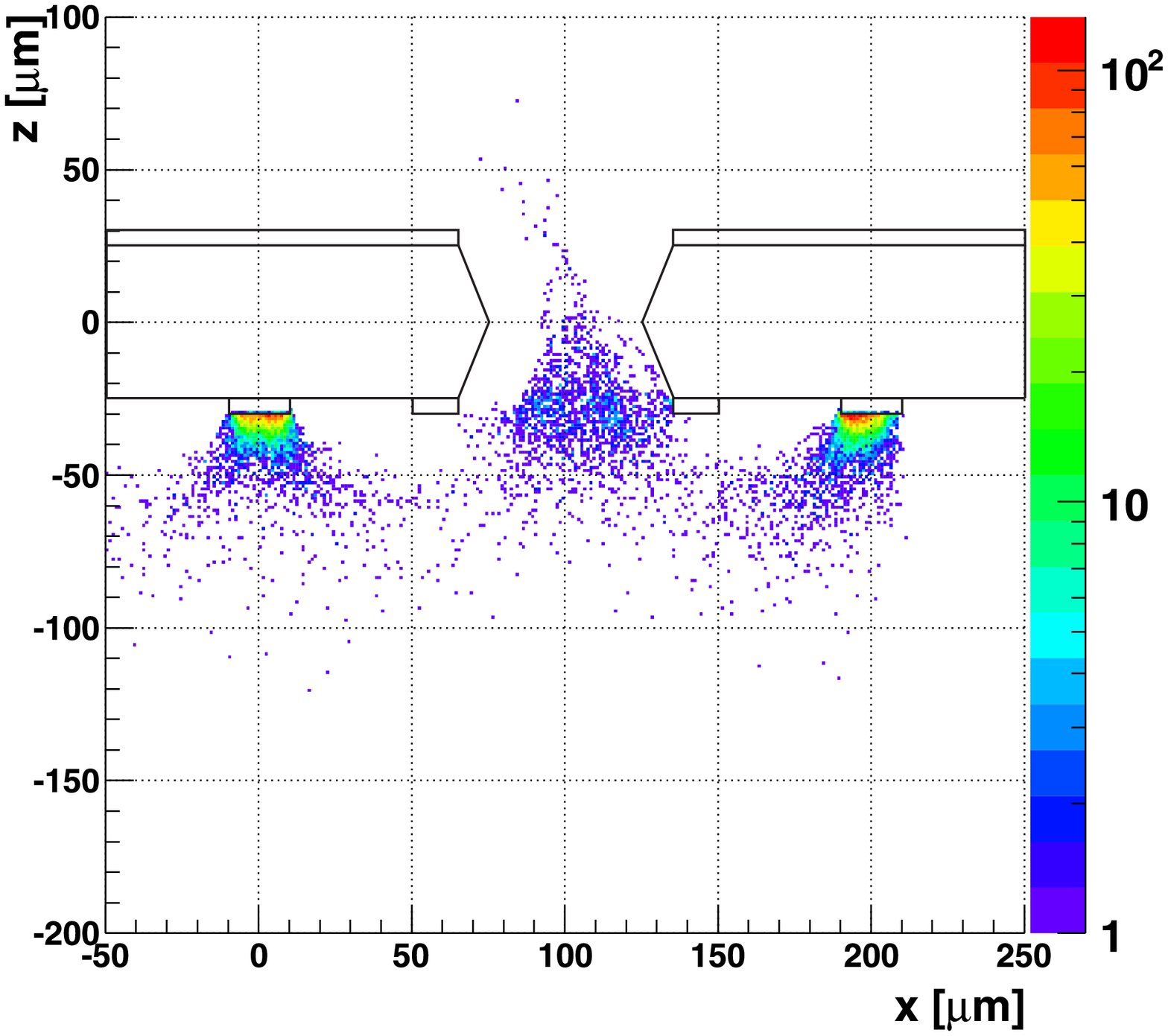}
\caption{Distribution of the $\left(x,z\right)$ coordinates of the points where excitations occurred, shown for $V_\mathrm{CT}=450$ V and $V_\mathrm{AC}=250$ V. The distribution refers to 2 primary electrons undergoing multiplication in a MHSP working with xenon at 1.0 bar and 300 K. The edges of the microstructure are proportionally represented at the same plot.}
\label{fig:mhsp_hxz}
\end{figure}

Figure~\ref{fig:mhsp_hxz} shows the distribution of the $\left(x,z\right)$ coordinates of the points where excitations occurred, for $V_\mathrm{CT}=450$ V and $V_\mathrm{AC}=250$ V (the maximum voltages considered in the simulations). The two regions with higher intensity reflect the two charge multiplication stages, one in the exit of the hole and the other in the vicinity of the anode strips. The presence of these two amplification stages is one of the figures of merit of the MHSP microstructure.

\section{Conclusions}

In this work, we estimated the total VUV EL yield of the GEM and MHSP microstructures working in pure xenon, through simulations at the microscopic level performed using a flexible C++ toolkit based on the Garfield and Magboltz programs. 

We demonstrated that the solid angle subtended by a photosensor placed below the GEM depends on the operating conditions. For the same pressure, as the voltage applied across the holes increases, the solid angle increases. For a fixed voltage, as the xenon pressure increases, the solid angle decreases. After correction for the solid angle variation, we were able to approximately reproduce the experimental measurements. 

The results obtained for the case of the MHSP are not consistent with measurements, although the typical two exponential amplification regimes are present in the Monte Carlo results. However, the comparison between the experimental yields for GEM and MHSP, when the voltage between the anode and the cathode of the latter is null, are unexpectedly very different. In these conditions, since the hole and induction field geometries are very similar, the MHSP should give similar response to that of the GEM. The disagreement suggests an experimental underestimation of the yield, supported by the used gas purification method: diffusion rather circulation.

\acknowledgments
This work was performed within the RD51 collaboration - Development of Micro-Pattern Gas Detectors Technologies. C. A. B. Oliveira was supported by the portuguese PhD fellowship with reference SFRH/BD/36562/2007. This work was supported by the projects PTDC/FIS/113005/2009 and CERN/FP/123604/2011 under the COMPETE and FCT (Lisbon) programs. This work was also supported by the Director, Office of Science, Office of Basic Energy Sciences of the US Department of Energy under Contract N. DEAC02-05CH11231.


\begin{thebibliography}{99}

\bibitem{xenon100}
The XENON100 Collaboration, \emph{First Dark Matter Results from the XENON100 Experiment}, {Physical Review Letters {\bf 105} (2010) 131302}.

\bibitem{arDM}
A Marchionni et al, \emph{ArDM: a ton-scale LAr detector for direct Dark Matter searches}, {Journal of Physics: Conference Series {\bf 308} (2011) 012006}.

\bibitem{Benetti_warp}
P. Benetti et al, \emph{First results from a dark matter search with liquid argon at 87 K in the Gran Sasso underground laboratory}, {Astroparticle Physics {\bf 28} (2008) 495}.

\bibitem{zeplin}
D.Yu. Akimov et al, \emph{WIMP-nucleon cross-section results from the second science run of ZEPLIN-III}, {Physics Letters B {\bf 709} (2012) 14}.

\bibitem{lux}
D. N. McKinsey et al, \emph{The LUX Dark Matter Search}, {Journal of Physics: Conference Series {\bf 203} (2010) 012026}.

\bibitem{exo}
EXO collaboration, \emph{Observation of Two-Neutrino Double-Beta Decay in Xe-136 with EXO-200}, {Physical Review Letters {\bf 107} (2011) 212501}.

\bibitem{NEXT_CDR}
The NEXT collaboration, \emph{The NEXT-100 experiment for neutrinoless double beta decay searches (Conceptual Design Report)}, {arXiv:1106.3630v1 [physics.ins-det]}.

\bibitem{NEXT_TDR}
The NEXT collaboration, \emph{NEXT-100 Technical Design Report (TDR). Executive Summary}, {Journal of Instrumentation {\bf 7} (2012) T06001}.

\bibitem{Tanaka}
Y. Tanaka, \emph{Continuous Emission Spectra of Rare Gases in the Vacuum Ultraviolet Region},
{Journal of the Optical Society of America {\bf 45} (1955) 710}.

\bibitem{Suzuki}
M. Suzuki, S. Kubota,
\emph{Mechanism of proportional scintillation in Argon, Krypton and Xenon},
{Nuclear Instruments \& Methods {\bf 164} (1979) 197}.

\bibitem{Oliveira_thesis}
C. A. B. Oliveira, \emph{Monte Carlo study of electroluminescence in gaseous detectors},
{PhD thesis, University of Aveiro, Portugal (2011) https://ria.ua.pt/bitstream/10773/7346/1/PhD\_thesis\_COliveira.pdf}

\bibitem{DARWIN}
L. Baudis, \emph{DARWIN dark matter WIMP search with noble liquids}, {arXiv:1201.2402v1 [astro-ph.IM], 2012} .

\bibitem{THGEM_DM}
M. Gai et al, \emph{Toward Application of a Thick Gas Electron Multiplier (THGEM) Readout for a Dark Matter Detector}, {arXiv:0706.1106v1 [physics.ins-det], 2007}.

\bibitem{GEM_dualphase}
F. Balau et al, \emph{GEM operation in double-phase xenon}, {Nuclear Instruments and Methods A {\bf 598} (2009) 126}.

\bibitem{Buzu_advances}
A. Buzulutskov, \emph{Advances in Cryogenic Avalanche Detectors}, {Journal of Instrumentation {\bf 7} (2012) C02025}.

\bibitem{cristina_GEM_MHSP}
C. M. B. Monteiro et al, \emph{Secondary scintillation yield from gaseous micropattern electron multipliers in direct Dark Matter detection}, {Physics Letters B {\bf 677} (2009) 133}.

\bibitem{apd_vuv_1}
P. K. Lightfoot et al, \emph{Optical readout tracking detector concept using secondary scintillation from liquid argon generated by a thick gas electron multiplier}, {Journal of Instrumentation {\bf 4} (2009) P04002}.

\bibitem{apd_vuv_2}
D. Yu. Akimov et al, \emph{Detection of scintillation light in liquid xenon by multipixel avalanche Geiger photodiode and wavelength shifter}, {Journal of Instrumentation {\bf 5} (2010) P04007}.

\bibitem{apd_radiopure}
T. Lux et al, \emph{Characterization of the Hamamatsu S8664 Avalanche Photodiode for X-Ray and VUV-light detection}, {arXiv:1108.5143v3 [physics.ins-det], 2012}.

\bibitem{cristina_GEM_THGEM}
C. M. B. Monteiro et al, \emph{Secondary Scintillation Yield from GEM and THGEM Gaseous Electron Multipliers for direct Dark Matter search}, {Physics Letters B {\bf 714} (2012) 18}.

\bibitem{OliveirauE}
C. A. B. Oliveira et al, \emph{A simulation toolkit for electroluminescence assessment in rare event experiments}, {Physics Letters B {\bf 703} (2011) 217}.

\bibitem{garfield}
\emph{Garfield++  --  simulation of tracking detectors}, {http://cern.ch/garfieldpp [2012, June 6$^{th}$]}

\bibitem{magboltz1}
S.F. Biagi,
\emph{Monte Carlo simulation of electron drift and diffusion in counting gases under the influence of electric and magnetic fields},
{Nuclear Instruments and Methods A {\bf 421} (1999) 234}.

\bibitem{magboltz2}
\emph{Magboltz - transport of electrons in gas mixtures}, {http://cern.ch/magboltz [2012, June 6$^{th}$]}

\bibitem{3d}
F. P. Santos et al,
\emph{Three-dimensional Monte Carlo calculation of the VUV electroluminescence and other electron transport parameters in xenon},
{Journal of Physics D {\bf 27} (1994) 42}.

\bibitem{expy_xe_ue}
C. M. B. Monteiro, L. M. P. Fernandes, J. A. M. Lopes, L. C. C. Coelho, J. F. C. A. Veloso, J. M. F. dos Santos, K. Giboni and E. Aprile,
\emph{Secondary scintillation yield in pure xenon},
{Journal of Instrumentation {\bf 2} (2007) P05001}.

\bibitem{expy_ar_ue}
C. M. B. Monteiro, J. A. M. Lopes, J. F. C. A. Veloso and J. M. F. dos Santos,
\emph{Secondary scintillation yield in pure argon},
{Physics Letters B {\bf 668} (2008) 167}.

\bibitem{ansys}
\emph{ANSYS}, {http://www.ansys.com/ [2012, June 6$^{th}$]}

\bibitem{fem}
Singiresu S. RAO, \emph{The Finite Element Method in Engineering}, {5$^{th}$ Edition}, {Elsevier, 2011}.

\end{thebibliography}
\end{document}